\documentclass[reprint,amsmath,amssymb,superscriptaddress,aps,prd]{revtex4-1}

\usepackage{graphicx}
\usepackage{aas_macros}
\usepackage{float}
\usepackage{amsmath}
\usepackage{appendix}
\usepackage[usenames, dvipsnames]{color}

\newcommand{\orcid}[1]{\href{https://orcid.org/#1}{\includegraphics[width=8pt]{orcid.png}}}

\newcommand{\degree}{\ensuremath{^{\circ}}}

\begin{document}

\title{Constraining Unmodeled Physics with Compact Binary Mergers from GWTC-1}

\author{Bruce Edelman}
    \email{bedelman@uoregon.edu}
\affiliation{Department of Physics, University of Oregon, Eugene, OR 97403, USA}
\author{F. J. Rivera-Paleo}
\affiliation{Departament d'Astronomia i Astrofísica, Universitat de València, 46100, Burjassot, Spain}
\author{J.D. Merritt}
\affiliation{Department of Physics, University of Oregon, Eugene, OR 97403, USA}
\author{Ben Farr}
\affiliation{Department of Physics, University of Oregon, Eugene, OR 97403, USA}
\author{Zoheyr Doctor}
\affiliation{Department of Physics, University of Oregon, Eugene, OR 97403, USA}
\author{Jeandrew Brink}
\affiliation{Department of Mathematics and Applied Mathematics, University of the Free State,
PO Box 339, Bloemfontein, South Africa}
\author{Will M. Farr}
\affiliation{Department of Physics and Astronomy, Stony Brook University, Stony Brook, NY 11794, USA}
\affiliation{Center for Computational Astrophysics, Flatiron Institute, New York NY 10010, USA}
\author{Jonathan Gair}
\affiliation{Max Planck Institute for Gravitational Physics (Albert Einstein Institute),
Am M{\"u}hlenberg 1, D-14476 Potsdam-Golm, Germany}
\author{Joey Shapiro Key}
\affiliation{University of Washington Bothell, 18115 Campus Way NE, Bothell, WA 98011, USA}
\author{Jess McIver}
\affiliation{University of British Columbia, Vancouver, BC V6T 1Z4, Canada}
\author{Alex B. Nielsen}
\affiliation{Max-Planck-Institut f{\"u}r Gravitationsphysik (Albert-Einstein-Institut), D-30167 Hannover, Germany}
\affiliation{Department of Mathematics and Physics, University of Stavanger, 4036 Stavanger, Norway}

\date{\today}

\begin{abstract}
We present a flexible model to describe the effects of generic deviations of observed gravitational wave signals from modeled waveforms in the LIGO and Virgo gravitational wave detectors. With the detection of 11 gravitational wave events from the GWTC-1 catalog, we are able to constrain possible deviations from our modeled waveforms. In this paper we present our coherent spline model that describes the deviations, then choose to validate our model on an example phenomenological and astrophysically motivated departure in waveforms based on extreme spontaneous scalarization. We find that the model is capable of recovering the simulated deviations. By performing model comparisons we observe that the spline model effectively describes the simulated departures better than a normal compact binary coalescence (CBC) model. We analyze the entire GWTC-1 catalog of events with our model and compare it to a normal CBC model, finding that there are no significant departures from the modeled template gravitational waveforms used.
\end{abstract}

\maketitle

\section{\label{sec:intro}Introduction}

General relativity (GR) has passed a multitude of tests over the past years
\cite{Will_2014}, but until the detection of gravitational waves (GWs) from
binary black holes\cite{GW150914_discovery,GW150914_PE} it had not been widely
tested for strong dynamical gravitational fields. Gravitational-wave astronomy
and more specifically that of compact binary coalescences (CBC) gives us access
to a genuinely strong gravitational field regime to both test GR
\cite{GW150914_testingGR, GW170817_testingGR, catalog_testingGR,
golden_bbh_testingGR}, and to provide constraints on new physics not predicted by GR modeled waveforms of these systems. Contemporary GW analyses employ matched filtering and
forward modeling techniques, which both inherently rely on accurately
modeled waveforms \cite{Moore_etal_2014, Smith_etal_2013, GW150914_PE,
LALinference}.  We introduce here a model that can account for and measure in
data deviations in phase and amplitude from a modeled waveform, either due to
approximations inherent in the waveform calculation or a mismatch between theory
(GR) and nature.

We present a parameterization that quantifies the deviations between the
observed waveform $(h_{\mathrm{intrinsic}})$  and the GR waveform models
$(h_{\mathrm{model}})$, with few assumptions about the deviations. This provides
the ability to perform additional tests of GR and also presents a generic model
for describing and possibly constraining additional effects in a binary merger like
presence of higher-order modes \cite{Chatziioannou_2019} and tidal effects
\cite{GW170817_tides}. Quantifying such deviations is one of the major challenges 
in GW data analysis. The numerical method we use to
parameterize deviations here is based on cubic spline interpolation in which the
deviations (in phase and amplitude) are modeled as independent cubic spline
functions interpolated from node points in the frequency domain. The cubic
spline interpolant generates deviations that vary smoothly in frequency, but
otherwise does not constrain the type or nature of deviation.

The splines employed to characterize the deviation provide a uniform way of describing GR departures rather than fitting separate parameters to the inspiral,  merger,  and ringdown (IMR) of the waveform separately as commonly done in GR IMR consistency tests \cite{IMR_testingGR, golden_bbh_testingGR, IMR_consistency_GWTC-1}. In addition, IMR consistency tests have the same limitation as matched filtering, as they inherently assume perfect accuracy of the template waveform.

Another common test of GR is the parameterized test where one expands the waveform model in different regimes with post-Newtonian (PN) correction parameters added \cite{Li:2011cg}. This test also builds in assumptions about how possible deviations may occur and has to fit different parts of the waveform with separate models as in the IMR consistency test. One commonly used class of model-agnostic tests of GR are residual \cite{catalog_testingGR} tests, where best-fit waveform models are subtracted from the data and normality tests (e.g., Anderson-Darling) are conducted on the residuals. Such tests would be sensitive to very large deviations from the signal model, but not to small but correlated deviations across many frequency bins. Our proposed model differs from these tests and constraints by allowing uncertainty in the template waveforms and letting it vary smoothly across the entire frequency range. Our model is able to describe and fit the inaccuracies in waveform models. With the assumption that the template GR waveform is completely accurate, it provides a clear way of describing and constraining unmodeled physics of our waveform models or departures from GR across the frequency range of the waveform.

In Section \ref{sec:spline} we describe the model and methods for incorporating it in the \texttt{LALInference} Bayesian Analysis software \cite{LALinference}.  Our implementation is similar to the calibration spline model described in \cite{B_Farr_etal_2014}. We then present simulated deviations on which we validated the performance of our model in Section \ref{sec:simulations}, followed by discussions and implications of the results. In Section \ref{sec:results} we present results of this model on the entire first LIGO and Virgo Gravitational Wave Transient Catalog (GWTC-1) \cite{catlog_paper} which includes the results of the first and second observing runs of the Advanced LIGO \cite{Adv-LIGO} and Virgo \cite{adv_VIRGO} detectors. This catalog of gravitational wave events includes ten binary black hole detections and one binary neutron star detection \cite{catlog_paper, GW170817_disc}. Lastly the results and conclusions are summarized and discussed in Section \ref{sec:conclusions}.

\section{\label{sec:spline}Methods}

\subsection{Waveform Representation}

When a gravitational wave enters a gravitational wave detector, the detector
records a data stream which we can describe in the frequency domain as $d(f) =
h_\mathrm{observed}(f) + n (f)$, which is an additive combination of a waveform
$h_\mathrm{observed}(f)$ and noise $n (f)$. The observed waveform in a detector
can be represented as the sum of the intrinsic waveform polarizations, projected
across that detector. This is done by multiplying the two (time-dependent)
antenna pattern terms of that detector, $F_{+}$ and $F_{\times}$, to the plus
and cross gravitational-wave polarizations as:
\begin{equation} \label{added}
    h_\mathrm{observed}(f) = h_{+}(f)F_{+}(f) + h_{\times}(f)F_{\times}(f)
\end{equation}
with
\begin{equation} \label{eqn2}
	h_{\text{+},\times}(f) = h_{\mathrm{model},+,\times}(f)\left[1 + \delta A(f)\right]\exp\left[i\delta\phi(f)\right].
\end{equation}

Since we are searching for deviations from the coherent modeled waveforms, every
detector observing a GW should see the same deviations. We model the
uncertainties in the intrinsic waveform, $h_\mathrm{model, +, \times}(f)$, as
frequency-dependent amplitude and phase departures in $h_\mathrm{+, \times}$
with respect to $h_\mathrm{model, +, \times}$. This is the same technique that
\citet{B_Farr_etal_2014} use to model calibration errors in each detector
independently. We also take the assumption that there are only two polarizations
and that GWs travel at the speed of light. While our model is degenerate with
the calibration model, coherent deviations observed across all detectors are
modeled with a single set of parameters for this spline model, as opposed to
modeling them independently in each detector using the calibration model. Thus
we expect our prior distribution to result in deviations to the intrinsic
waveform being preferably described by the coherent spline, and the calibration
splines to measure only detector-dependent deviations.

We assume that phase deviations are small and under this assumption, we can
approach the exponential term as
\begin{equation}
\exp\left(i \delta \phi(f) \right) = \frac{2+i \delta\phi(f)}{2-i \delta\phi(f)} +{\cal{O}} (\delta \phi^3)
\end{equation}
 which is more computationally efficient \cite{B_Farr_etal_2014}. Then, we can
 rewrite the intrinsic waveform as:
\begin{equation} \label{eqn3}
    h_{\text{+},\times}(f) =  h_{\mathrm{model},+,\times}(f)\left[1 +\delta A(f)\right]\frac{2+i\delta\phi(f)}{2-i\delta\phi(f)}
\end{equation}
This replacement agrees with the exponential term to third order for small phase
deviations, and differs by 5\% from the exponential term for the largest
simulated deviation used in this paper of 60 degrees.

The intrinsic waveform, after being modified as Eq.\ \ref{eqn3}, is then
projected across the detectors to get $h_\mathrm{observed}$ in each detector as
in Eq.\ \ref{added}. Despite the expectation that these departures are
small, they have the potential to impact the measurement of all parameters of
the source (masses, spins, distance, etc.) \cite{Smith_etal_2013}. A consequence
of modeling these deviations with a purely phenomenological model is that we can
no longer trust the inference of astrophysically modeled parameters when
deviations are present.

Under the assumption that $\delta A(f)$ and $\delta\phi(f)$ vary smoothly in
frequency, they can be modeled by a spline function \cite{B_Farr_etal_2014}.

\subsection{Spline Model}

A spline function is a piece-wise polynomial interpolation that obeys smoothness
conditions at the nodal points where the pieces connect. In the following, we
use the case of cubic splines defined by $15$ nodal points confined to a finite
frequency interval. Formally these departures can be written as
\begin{eqnarray}
  \delta A(f) &=& \mathcal{I}_3(f;\{f_i,\delta A_i\}), \label{eqn4}\\
  \delta \phi(f) &=& \mathcal{I}_3(f;\{f_i,\delta \phi_i\}), \label{eqn5}
\end{eqnarray}
where $\mathcal{I}_3$ is a cubic spline interpolant, the $\{f_i\}$ are the nodes
of the spline interpolant in frequency, and $\{\delta A_i\}$ and $\{\delta
\phi_i\}$ are the values of the spline at those nodal points. To better
generalize the model to fit a larger variety of possible departures, we freely
let the nodal points move around in frequency space during sampling (with the
condition that they do not exchange orders or get too close to each other) after
being initialized linearly in log-frequency space, as done in
\citet{Vitale_etal_2012,B_Farr_etal_2014}. We choose the node locations
to be the same for the amplitude and phase spline functions as we expect deviations to happen
at similar frequencies and to reduce our model degrees of freedom. One could choose an
independent set of node locations for phase and amplitude if expecting to have deviation 
effects that alter the amplitude and phase at different points in frequency. We prevent 
nodal points from getting arbitrarily close as this causes the spline to be too 
sensitive to very small changes in node positions, creating extreme deviations 
to satisfy the smoothness conditions of the spline function. The parameters added 
to our inference for this model are then the $\{\delta A_{i}\}$, $\{\delta \phi_{i}\}$ 
and $\{f_{i}\}$.

With our assignment of Gaussian priors (see \S\ \ref{sec:statistical-framework})
to the $\delta A$ and $\delta \phi$ parameters, our spline model implements a
Gaussian process prior for the waveform deviations from the model (c.f.\
\cite{Kimeldorf1970}).

\subsection{Statistical Framework}
\label{sec:statistical-framework}

We place a Gaussian prior on the departure parameters, $\{\delta A_{i}\}$ and
$\{\delta \phi_{i}\}$, centered around zero, with $\sigma_A$ and $\sigma_\phi$
characterizing our prior uncertainties about the magnitude of the departures in
amplitude and phase from the modeled waveforms.
\begin{eqnarray}
  p(\delta A_i) &=& N(0,\sigma_A) \label{eqn6} \\
  p(\delta \phi_i) &=& N(0,\sigma_\phi) \label{eqn7} \\
  p(f_\mathrm{i}) &=& U(f_\mathrm{low}, f_\mathrm{high}) \label{fnode_prior}
\end{eqnarray}
It is important however, not to think of the Gaussian priors on the node values
of $\{\delta A_{i}\}$ and $\{\delta \phi_{i}\}$ as the broadband uncertainties
of the interpolated spline function across the frequency range. In practice the
prior widths on the nodes are wider than the broadband uncertainty resulting
from sampling the prior.  We constrain the spline nodes to be increasing in
frequency:
\begin{equation} \label{eqn8}
    f_\mathrm{i-1} < f_\mathrm{i} < f_\mathrm{i+1}.
\end{equation}

The spline model introduces some challenges that need to be accounted for, the
first being the freedom in the frequency values of the node points. Since we
also use the node positions in frequency space as sampling parameters and we
place uniform priors on the spline node locations across the frequency band from
$f_\mathrm{low}$ to the Nyquist frequency, the model is degenerate under
exchange of node positions. To circumvent this degeneracy, we impose that the
node positions stay ordered as in Eq.\ \ref{eqn8}. These parameters can then
be fit and the corresponding calibration errors marginalized over during
inference.

Another challenge with the spline function is that if two nodes get too close
together, obeying the conditions required of the cubic spline can lead to the
interpolated deviations becoming very extreme. To account for this we prevent
the frequency nodes from getting closer than 4 frequency bins away from one
another as shown in equation \ref{eqn9}, where the frequency bin width (in Hz)
is determined from the segment length of data we are analyzing as $df = 1 / T$
with T the segment length.   That is, we reject any configurations with

\begin{equation} \label{eqn9}
   f_\mathrm{i+1} - f_\mathrm{i} < 4 * \delta f = \frac{4}{T_\mathrm{obs}}.
\end{equation}

Other spline based interpolation methods such as \texttt{BayesLine} combat this by keeping 
the spline nodes on a fixed frequency grid then turns them on/off during inference 
using trans-dimensional Markov-Chain Monte Carlo sampling \cite{Littenberg_2015}. 
We experimented with a few other fixes to this issue. We implemented a Gaussian
prior on every frequency bin location; we also tried evaluating the Gaussian
prior on some number of points between nodes of the spline interpolants to
disfavor the large spline excursions as well. In practice both of these priors
turned out to be too restrictive while attempting to recover the simulated
deviations presented in Section \ref{sec:simulations}; they may still be useful
when exploring astrophysical events. 

\begin{figure}[h]
  \centering
  \includegraphics[width=0.4\textwidth]{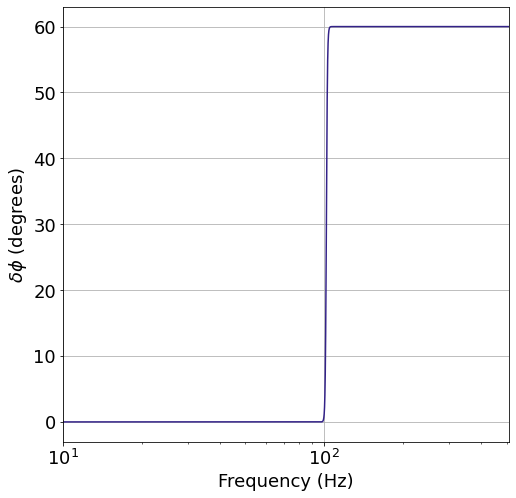}
  \caption{Simulated phase deviation $\delta \phi$ for extreme spontaneous scalaraization toy model.}
\label{fig:phi}
\end{figure}

\begin{figure}[h]
  \centering
  \includegraphics[width=0.4\textwidth]{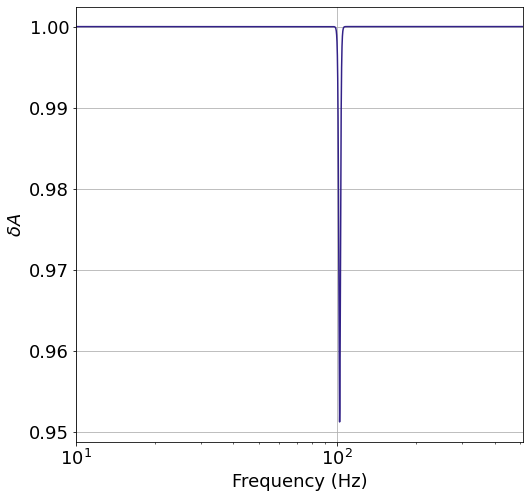}
  \caption{Simulated amplitude deviation $\delta A$ for extreme spontaneous scalaraization toy model.}
\label{fig:A}
\end{figure}

The last challenge is that our model of amplitude deviations is perfectly
degenerate with the distance to the signal. The distance to the source can be
increased while simultaneously producing a positive amplitude deviation across
the entire frequency band of the signal to compensate. The default astrophysical
prior used in \texttt{LALInference} \cite{LALinference} is $\propto D_{L}^{2}$ with
$D_{L}$ the luminosity distance. When this prior distribution is coupled with
the zero-mean Gaussian priors on amplitude deviations, the strong $\propto
D_{L}^{2}$ prior on distance almost always results in systematically positive
amplitude deviations in the spline component of the model. However, by allowing
for broad-band changes to the amplitude of the signal in the first place we are
no longer able to meaningfully infer the distance to the source. In other words,
we use $D_{L}$ (with a uniform prior) as a phenomenological parameter to fit the
broad-band amplitude of the signal, and the spline model describes any
frequency-dependent deviations that may be present.

With the prior assumptions and modified waveform we can now construct the
posterior distribution according to Bayes' Theorem,
\begin{multline} \label{posterior}
    p(\boldsymbol{\theta}, \{\delta A_{i}\}, \{\delta \phi_{i}\}, \{f_{i}\} | d) \propto \\
    \mathcal{L}(d|\boldsymbol{\theta},\{\delta A_{i}\}, \{\delta \phi_{i}\}, \{f_{i}\})p(\boldsymbol{\theta})p(\{\delta A_{i}\})p(\{\delta \phi_{i}\})p(\{f_{i}\})
\end{multline}
then use the \texttt{LALInference} Markov Chain Monte Carlo (MCMC) algorithm to draw
samples from the posterior distribution in Eq.\ \ref{posterior}, with
$\boldsymbol{\theta}$ the normal CBC parameters, $d$ the gravitational wave
strain data, $\mathcal{L}$ the standard GW likelihood with the modified
intrinsic waveform as shown in Eq.\ \ref{eqn3}.

\section{\label{sec:simulations}Simulated Deviations}

In order to validate our spline model, we run on data with astrophysically motivated deviations included. We choose a toy model that presents an extreme case of spontaneous scalarization \cite{Sampson_et_al_2014} which modifies a simulated high-mass BBH waveform with similar astrophysical parameters as GW150914 \cite{GW150914_discovery, GW150914_PE}. We generate two waveforms where one has simulated modifications according to the toy model and the other is exactly as the modeled GR waveform describes. We then simulate colored Gaussian noise according to the the Advanced LIGO design sensitivity noise curve or power spectral density (PSD) \cite{Adv-LIGO} and add it to the simulated waveforms.

\begin{table}[h!]
\centering
\begin{tabular}{ |c|c| }
 \hline
 $m_{1}$ & 36 $M_{\odot}$ \\
 $m_{2}$ & 29 $M_{\odot}$ \\
 $D_\mathrm{L}$ & 450 Mpc \\
 $\phi$ & 2.76 rad \\
 $\alpha$ & 1.37 rad \\
 $\delta$ & -1.26 rad \\
 $S_\mathrm{1,z}$ & 0.0 \\
 $S_\mathrm{2,z}$ & 0.0 \\
 \hline
\end{tabular}
\caption{Parameters for simulated validation signals}
\label{table:1}
\end{table}

\begin{figure*}[t]
\centering
\includegraphics[width=.9\textwidth]{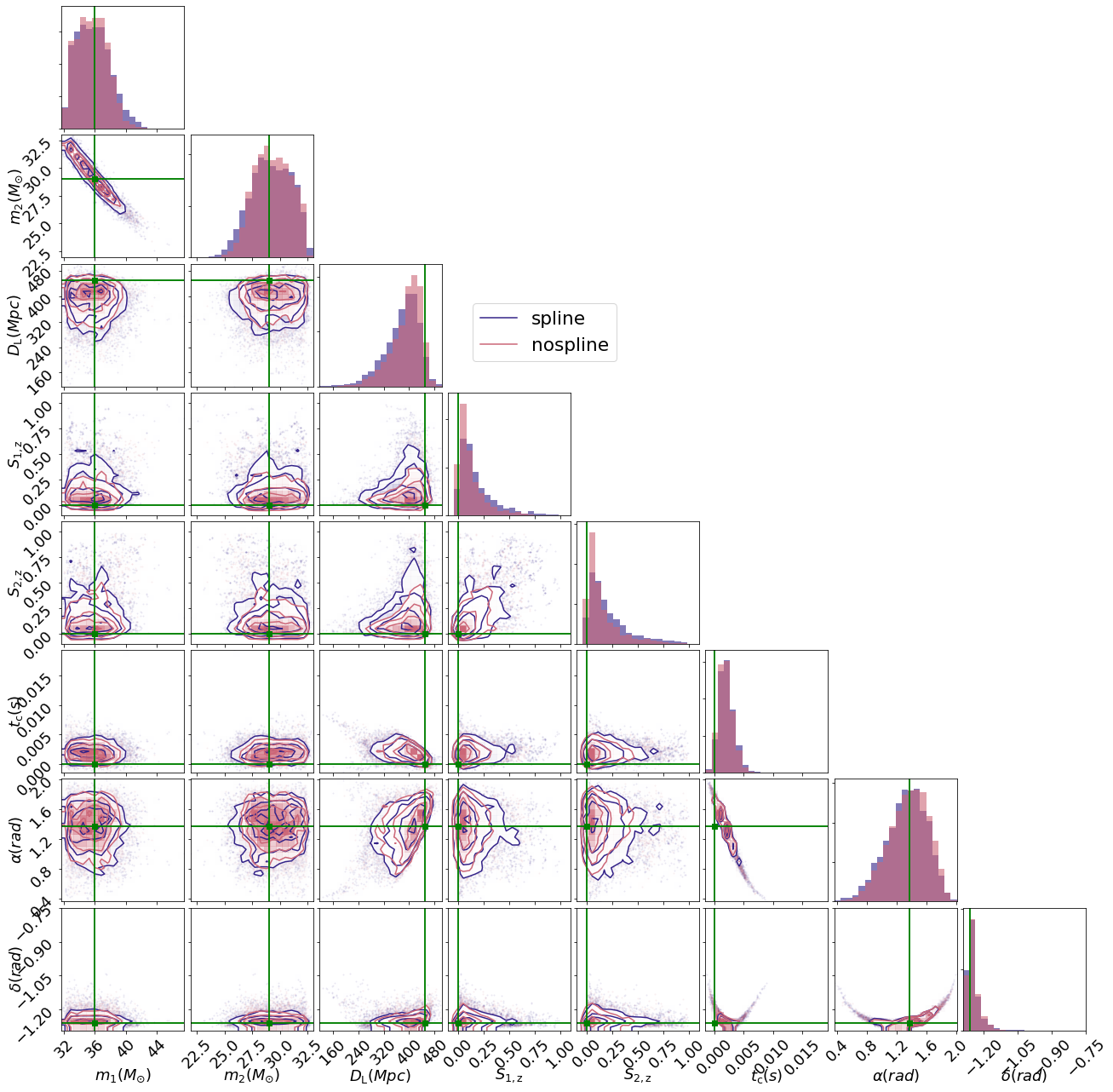}
\caption{Corner plots showing the 1-D and 2-D marginalised posterior distributions for simulated parameters for PE runs on the Unmodified Signal with spline model on (purple) or off (pink). This demonstrates the impact of the model flexibility on astrophysical parameter uncertainties.}
\label{fig:pure_corner}
\end{figure*}

\begin{figure*}[t]
\centering
\includegraphics[width=.9\textwidth]{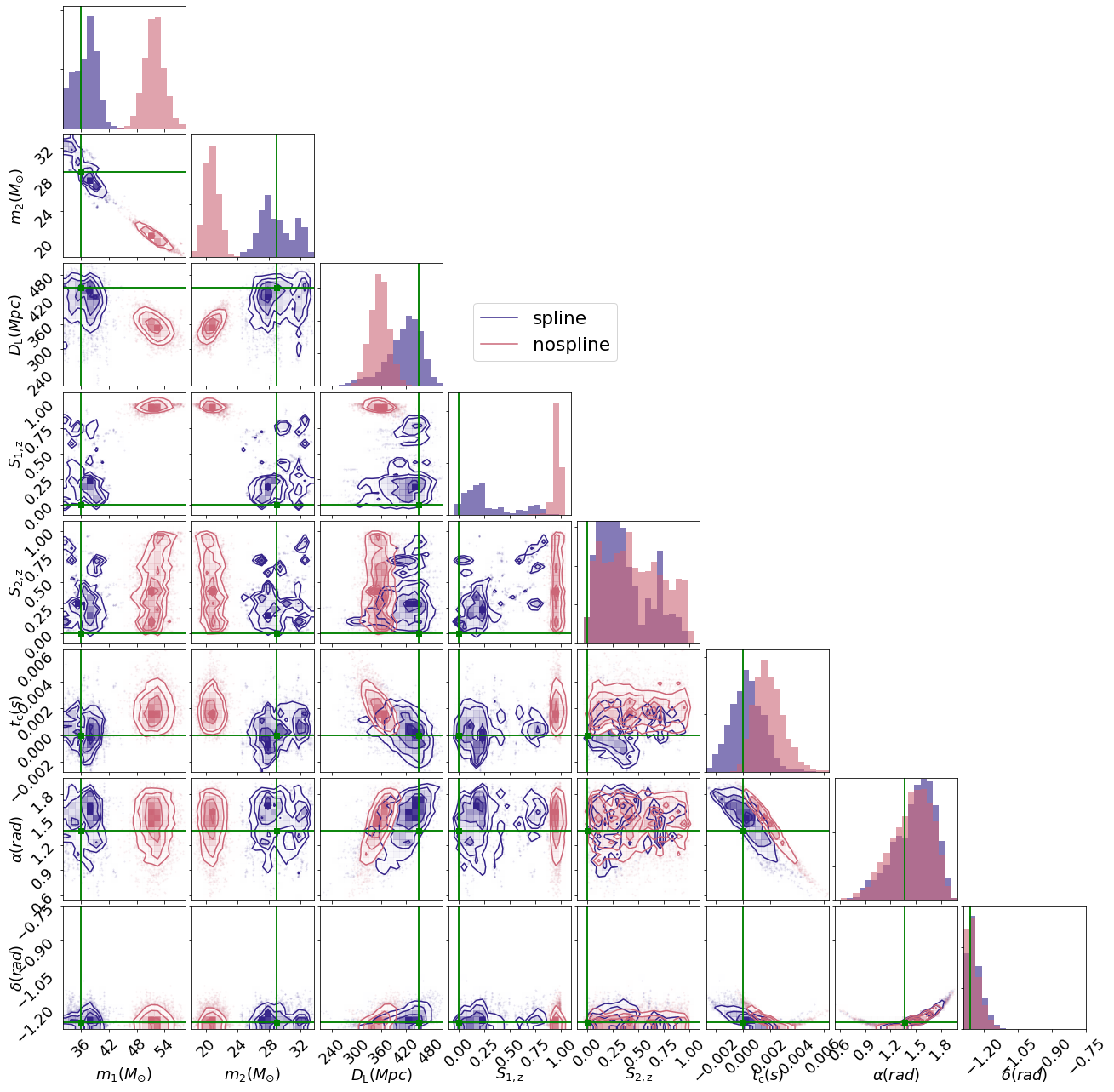}
\caption{Corner plots showing the 1-D and 2-D marginalised posterior distributions for simulated parameters for PE runs on the Modified Signal with spline model on (purple) or off (pink). This demonstrates inaccuracies in parameter estimation performed on signals containing deviations from the physically modeled waveforms.}
\label{fig:mod_corner}
\end{figure*}

We first generate a GR-based frequency domain waveform using the IMRPhenomD \cite{IMRPhenomPv2} waveform approximant with the simulated parameters shown in Table \ref{table:1}, then modify the same waveform according to our toy model described above extreme scalarization case \cite{Sampson_et_al_2014}. For the phase $\delta \phi$, the toy model includes an abrupt increase of 60 degrees centered at $f_{z} = 102 \text{Hz}$, with a width of $df = 1 \text{Hz}$. The amplitude temporarily drops by $5\%$ in frequency, again centered at $f_{z} = 102 \text{Hz}$ and with a width of $df = 1 \text{Hz}$. To get these modifications we used these parameters in Eqs.\ \ref{farr_mod} and \ref{farr_mod2} with $dA = 0.1$, $d\phi = 60 \degree$.

\begin{equation} \label{farr_mod}
            \begin{aligned}
            \delta A(f) = e^{\frac{1}{2}dA \Bigl(\tanh\left(\frac{f-f_{z}}{df}\right)^{2} -\tanh\left(\frac{f_\mathrm{ref}-f_{z}}{df}\right)^{2} \Bigr)}
            \end{aligned}
\end{equation}{}
\begin{equation} \label{farr_mod2}
            \begin{aligned}
                \delta \phi (f) = \frac{1}{2} d\phi \Bigl[\tanh\left(\frac{f-f_{z}}{df}\right) - \tanh\left(\frac{f_\mathrm{ref}-f_{z}}{df}\right)\Bigr]
            \end{aligned}{}
\end{equation}{}

\subsection{Results on simulated signals}

\begin{figure*}[t!]
\centering
\begin{tabular}{ll}
\includegraphics[width=.45\textwidth]{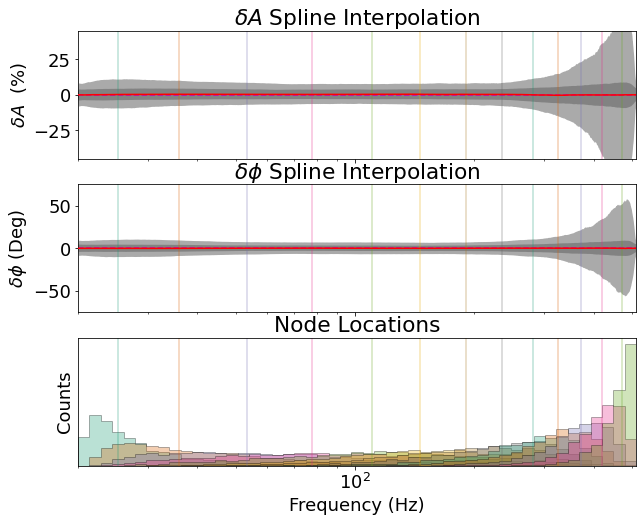}
&
\includegraphics[width=.45\textwidth]{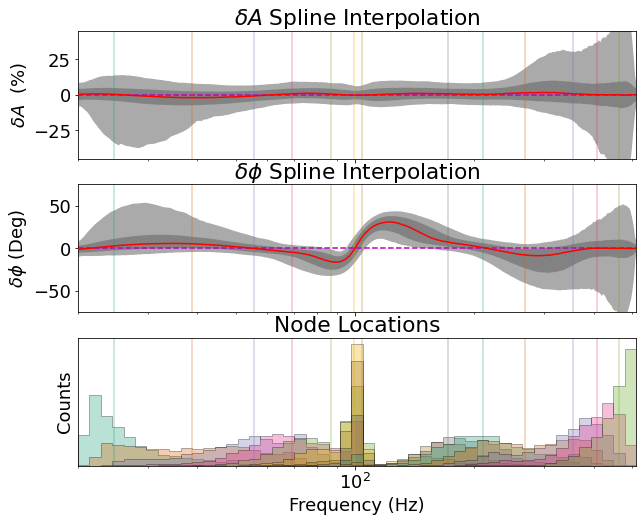}
\end{tabular}
\caption{Spline Interpolation of unmodified (left) and modified (right) signal deviations. 1 and 2 $\sigma$ credible intervals (grey) and the median spline (red) are shown with top panel the amplitude deviations and middle panel the phase deviations. In the bottom frame we plot the node position posterior distributions, which clump towards the frequency of deviations in the modified case in the bottom right most panel. For the unmodified case they are more uniformly distributed as they are exploring the prior.}
\label{fig:spline}
\end{figure*}

\begin{figure}[t]
    \centering
    \includegraphics[width=.5\textwidth]{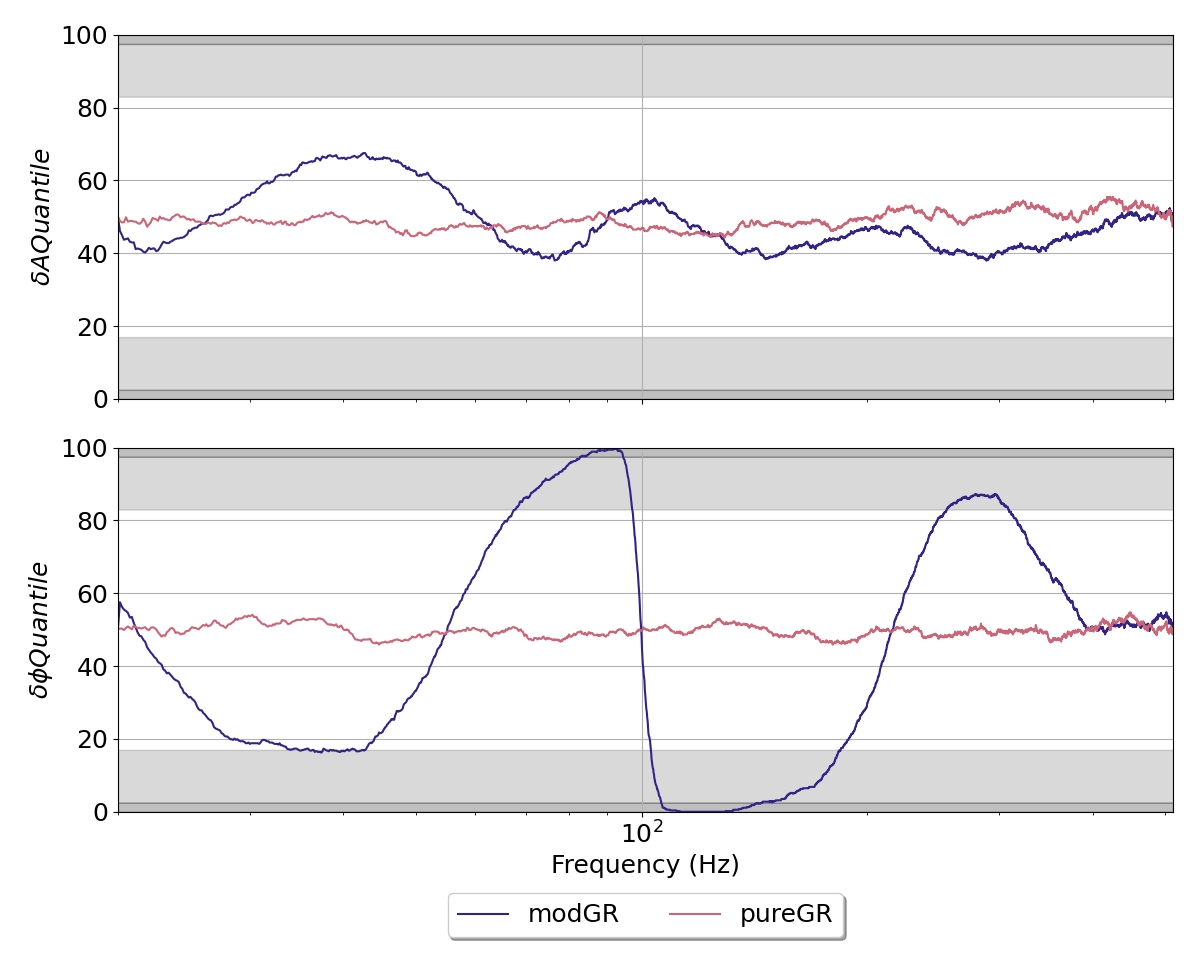}
    \caption{Posterior quantile of the spline interpolant that 0-deviation corresponds to for the simulated unmodified (pink) and modified (purple) signals.}
    \label{fig:quantiles}
\end{figure}{}

To compare the effect of the waveform modification and the spline model, we ran the \texttt{LALInference} parameter estimation software on both the modified and unmodified signals with our spline deviation model turned on and off. Figures \ref{fig:pure_corner} and \ref{fig:mod_corner} show the corner plots for the simulated intrinsic (component masses, spins, etc) and extrinsic (luminosity distance and sky localisation) GW parameters comparing the 1-D and 2-D marginalised posterior distributions with the spline model on or off. In figure \ref{fig:pure_corner} we see that there is minimal difference in posteriors for the unmodified signal for most of the parameters which is what we would expect for the case of no deviations. We also expect to see greater uncertainties on certain parameters with the spline model as there are possible degeneracies with the spline and other parameters as seen in the different 1-D spin parameter posteriors in figure \ref{fig:pure_corner}. This means the normal CBC model or template waveform is able to explain the entire coherent signal in the data.

For the Modified case (fig \ref{fig:mod_corner}) we see that the simulated values are included in most of the spline model posterior distributions but not the posteriors with the spline model inactive. This is because the modified signal includes the abrupt (in frequency) modifications or deviations simulated and these types of extreme abrupt deviations are very poorly described by the template waveform models and especially the IMRPhenomPv2 waveform template used in this analysis. However, we see that with our spline model turned on, the posterior distributions do encompass most of the simulated parameter values. This illustrates that parameter estimation (PE) without including deviation parameters would not be reliable at estimating the true signal parameters in cases where there may be departures from the waveform template used. In particular we see from figure \ref{fig:mod_corner} that the masses and trigger time posterior distributions are more consistent with the simulated values with the spline model fitting the deviations.

Figure \ref{fig:spline} shows the interpolated spline functions for the deviations in amplitude and phase, $\delta A$ and $\delta \phi$, for both the modified and unmodified simulated signals (right, left). The plots show the 1$\sigma$ and 2$\sigma$ bounds in the spline interpolants along with the median. We see here that our model consistently recovers zero deviations for the unmodified signal across the frequency band. The ranges included in the 1$\sigma$ and 2$\sigma$ bands show the exploration of the prior bounds while sampling while also being symmetric around the median. Looking at the modified case, we see a presence of deviations away from zero in phase around 100 Hz. The phase recovery does not show the clear step function behavior that was simulated and shown in figure \ref{fig:phi} as that extreme deviation is disfavored by the prior, because the priors used on the node positions are Gaussian distributions centered about zero; to recover the large flat step in frequencies greater than 100 Hz would require very low prior probability at those nodes. The spline posterior does however show that there is a transition at the modified frequency ($\sim100$ Hz) and shows that the phase increases roughly by 60 degrees as we simulated. The model compensates for this by ``ramping'' down the phase modification so that it slopes back to zero after the merger frequency ($\sim250$ Hz for this signal) since there is no signal content to infer from after merger. The model may also be able to coherently ramp down as a result of possible degeneracies with other parameters. The amplitude recovery is less revealing since the simulated deviation is on order of the prior width along with the sharp resolution of the deviation in frequency, it is harder to clearly recover that feature in our model. The presence of deviations is corroborated by the fact that the posterior distribution on the deviations excludes zero deviation at some frequencies at the 95\% level, which we do see in our modified case. To focus on this we can calculate the posterior quantile of phase and amplitude deviation that the x-axis (0-deviations) falls at for each frequency bin. This is shown in figure \ref{fig:quantiles}.

From this we see in the phase plot that there are significant portions of the frequency range where the phase interpolant ruled out zero deviations at greater than the 99\% credible level. The amplitude recovery shows less excursion which tells us we would not be able to constrain amplitude deviations of this type and magnitude. We do see significantly more freedom in amplitude than the no-modifications case which considerably constrained the amplitude interpolants to zero, however zero deviations lies within the $1\sigma$ range across the frequency band meaning it could still be consistent with no deviations present in amplitude.\break

\subsection{Model Comparisons}
\begin{figure}[t]
    \centering
    \includegraphics[width=.5\textwidth]{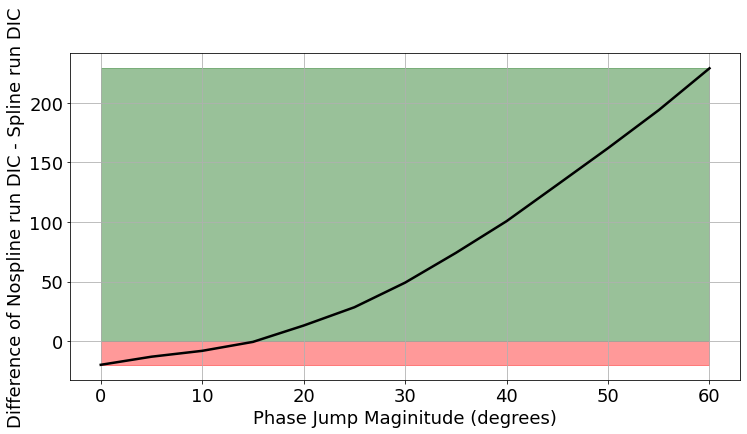}
    \caption{The DIC value from spline-model-off run minus spline-model-on run for different simulated phase deviations. Green shaded regions are where the spline model is preferred and red shaded region is where the spline model off is preferred.}
    \label{fig:DICramp}
\end{figure}

A useful way to see how well the spline model performs on a given segment of data is to perform a model comparison between the spline on and spline off models. To do this we take an Information Theory approach by computing the Deviance Information Criterion (DIC) \cite{DIC_12, DIC_2002}. This measure of fitness has the feature that lower values correspond to a better fit and includes an ``Occam's Factor'' that penalizes models with greater numbers of parameters. To effectively compare the spline on to spline off models we take the DIC value from the spline off and subtract the DIC value from the spline on model giving us a single value that is positive if the spline model is preferred and negative if the spline model is disfavored. We now perform a study where we run on simulated signals with modifications as before but with increasing jumps in phase from 0 to 60 degrees incrementing by 5 degrees. Looking at figure \ref{fig:DICramp}, we see that as the phase jump increases the spline model becomes more favored. We can compare the DIC differences from figure \ref{fig:DICramp} and to the differences for the un-modified signal which has DIC difference of -74.46, showing that for the DIC the spline model off is preferred at about the same level the spline model on is preferred for a phase jump of 35 degrees.

\subsection{Model Limitations and Alternative Parameterizations}

We have attempted a few other iterations of this model to increase its flexibility and performance on simulated signals which were not included in our final analyses. The first was to parameterize the deviations in the derivative of the phase/amplitude. This was tried for the same reason that the model was insensitive to the step function deviation that we have simulated in the phase. The step function was very incompatible with the normal model priors as each node after stepping up is penalized from our priors yet if we parameterized in the derivative the step function derivative looks like a delta function which is much more compatible with zero-centered Gaussian priors on the spline nodes. In practice this brought more challenges than it solved increasing other degeneracies with parameters and did not seem to qualitatively improve our efficiency or performance of correctly fitting the simulated deviations.

As a further test of our model, we attempted to recover modifications to a neutron star-black hole (NSBH) merger waveform. In the event of tidal disruption of the neutron star, there is an expected deviation to the waveform predicted by GR for non-deformable bodies. Specifically, we constructed a toy model featuring a roll-off in amplitude beyond a disruption frequency. Physically, this corresponds to a spreading and redistribution of mass after the moment of disruption, which would decrease the intensity of GWs emitted from then on \cite{NSBH_tidal_amp, NSBH-TIDAL}. We made no change to the phase. Our model was consistently unable to recover these deviations. This is likely due to more degeneracies in our parameters, and specifically in component mass. By moving to a higher mass, parameter estimation can push the amplitudes lower at high frequencies, and the additional flexibility of the spline model was unable to capture any of the deviation. Better results may be had with a more realistically modified NSBH waveform, or by changing the way we manage degeneracies in our parameters.

\section{\label{sec:results}Results from LIGO-Virgo Public Data}

\begin{table*}[t]
\centering
\begin{tabular}{ |c||c|c|c|c|c|c|c| }
 \hline
 \multicolumn{8}{|c|}{GWTC-1 Events} \\
 \hline \hline
Event & $\chi_{eff}$ & $d_{L}$ (Mpc) & $m_{1}$ ($M_{\odot}$) & $m_{2}$ ($M_{\odot}$) & $\mathcal{M}$ ($M_{\odot}$) & $\rho_\mathrm{gstlal}$ & Source \\
\hline
GW150914 & $-0.01_{-0.13}^{+0.12}$ & $430.0_{-170.0}^{+150.0}$ & $35.6_{-3.0}^{+4.8}$ & $30.6_{-4.4}^{+3.0}$ & $28.6_{-1.5}^{+1.6}$ & 24.4 & BBH \\
GW151012 & $0.04_{-0.19}^{+0.28}$ & $1060.0_{-480.0}^{+540.0}$ & $23.3_{-5.5}^{+14.0}$ & $13.6_{-4.8}^{+4.1}$ & $15.2_{-1.1}^{+2.0}$ & 10.0 & BBH \\
GW151226 & $0.18_{-0.12}^{+0.2}$ & $440.0_{-190.0}^{+180.0}$ & $13.7_{-3.2}^{+8.8}$ & $7.7_{-2.6}^{+2.2}$ & $8.9_{-0.3}^{+0.3}$ & 13.1 & BBH \\
GW170104 & $-0.04_{-0.2}^{+0.17}$ & $960.0_{-410.0}^{+430.0}$ & $31.0_{-5.6}^{+7.2}$ & $20.1_{-4.5}^{+4.9}$ & $21.5_{-1.7}^{+2.1}$ & 13.0 & BBH \\
GW170608 & $0.03_{-0.07}^{+0.19}$ & $320.0_{-110.0}^{+120.0}$ & $10.9_{-1.7}^{+5.3}$ & $7.6_{-2.1}^{+1.3}$ & $7.9_{-0.2}^{+0.2}$ & 14.9 & BBH \\
GW170729 & $0.36_{-0.25}^{+0.21}$ & $2750.0_{-1320.0}^{+1350.0}$ & $50.6_{-10.2}^{+16.6}$ & $34.3_{-10.1}^{+9.1}$ & $35.7_{-4.7}^{+6.5}$ & 10.8 & BBH \\
GW170809 & $0.07_{-0.16}^{+0.16}$ & $990.0_{-380.0}^{+320.0}$ & $35.2_{-6.0}^{+8.3}$ & $23.8_{-5.1}^{+5.2}$ & $25.0_{-1.6}^{+2.1}$ & 12.4 & BBH \\
GW170814 & $0.07_{-0.11}^{+0.12}$ & $580.0_{-210.0}^{+160.0}$ & $30.7_{-3.0}^{+5.7}$ & $25.3_{-4.1}^{+2.9}$ & $24.2_{-1.1}^{+1.4}$ & 15.9 & BBH \\
GW170817 & $0.0_{-0.01}^{+0.02}$ & $40.0_{-10.0}^{+10.0}$ & $1.46_{-0.1}^{+0.12}$ & $1.27_{-0.09}^{+0.09}$ & $1.186_{-0.001}^{+0.001}$ & 33.0 & BNS \\
GW170818 & $-0.09_{-0.21}^{+0.18}$ & $1020.0_{-360.0}^{+430.0}$ & $35.5_{-4.7}^{+7.5}$ & $26.8_{-5.2}^{+4.3}$ & $26.7_{-1.7}^{+2.1}$ & 11.3 & BBH \\
GW170823 & $0.08_{-0.22}^{+0.2}$ & $1850.0_{-840.0}^{+840.0}$ & $39.6_{-6.6}^{+10.0}$ & $29.4_{-7.1}^{+6.3}$ & $29.3_{-3.2}^{+4.2}$ & 11.5 & BBH \\
 \hline
\end{tabular}
\caption{Posterior estimates for LIGO-Virgo's GWTC-1 catalog of events with +/- 1$\sigma$ errors shown as well}
\label{table:catalog}
\end{table*}

\begin{figure}[t]
    \centering
    \includegraphics[width=0.5\textwidth]{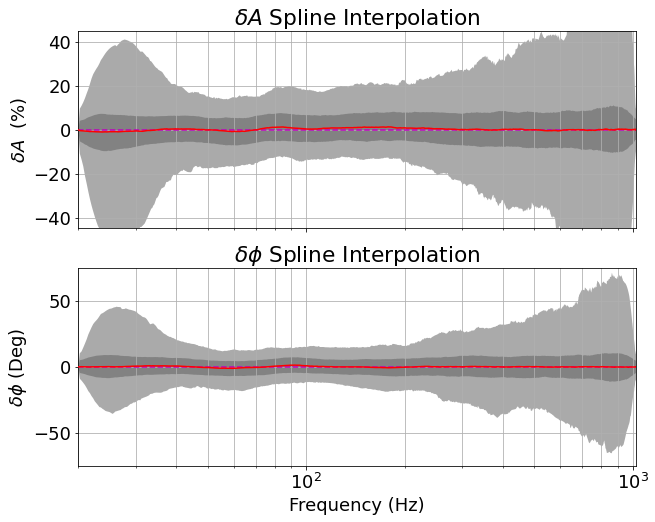}
    \caption{Spline interpolation of GW170823 with 1 and 2 $\sigma$ credible intervals (grey) and the median spline interpolant (red) shown.}
    \label{fig:spline_gw170823}
\end{figure}{}

\begin{figure}[h!]
    \centering
    \includegraphics[width=0.5\textwidth]{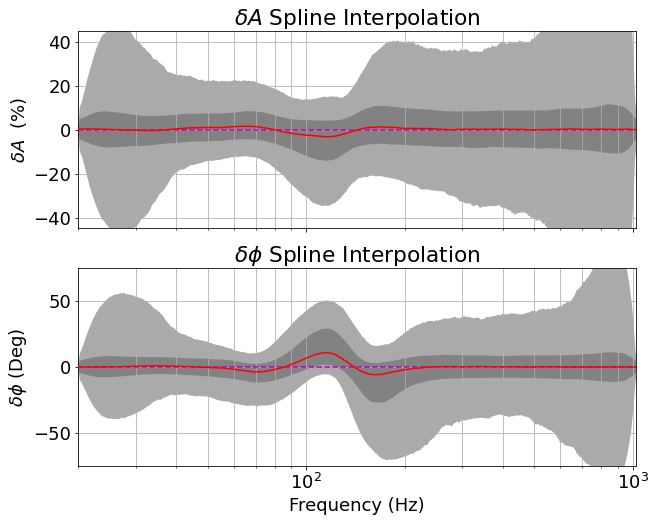}
    \caption{Spline interpolation of GW170729 with 1 and 2 $\sigma$ credible intervals (grey) and the median spline interpolant (red) shown.}
    \label{fig:spline_gw170729}
\end{figure}{}

\begin{figure}[t]
    \centering
    \includegraphics[width=0.5\textwidth]{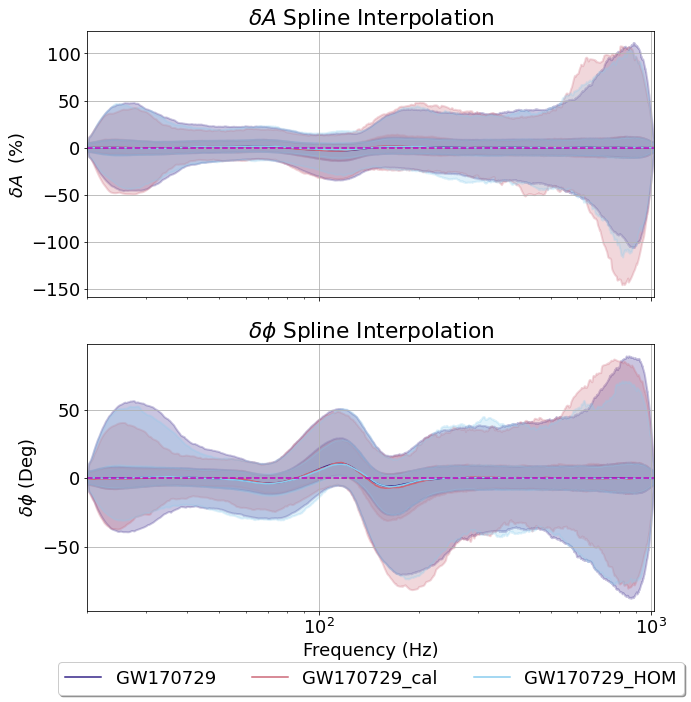}
    \caption{Spline interpolation of GW170729 with 1$\sigma$ percent credible intervals shown, comparing runs on GW170729 with calibration turned on/off and using a HOM waveform also with calibration turned off.}
    \label{fig:170729_cal_HOM}
\end{figure}{}
\begin{figure*}[t]
    \centering
    \includegraphics[width=0.9\textwidth, height=0.4\textheight]{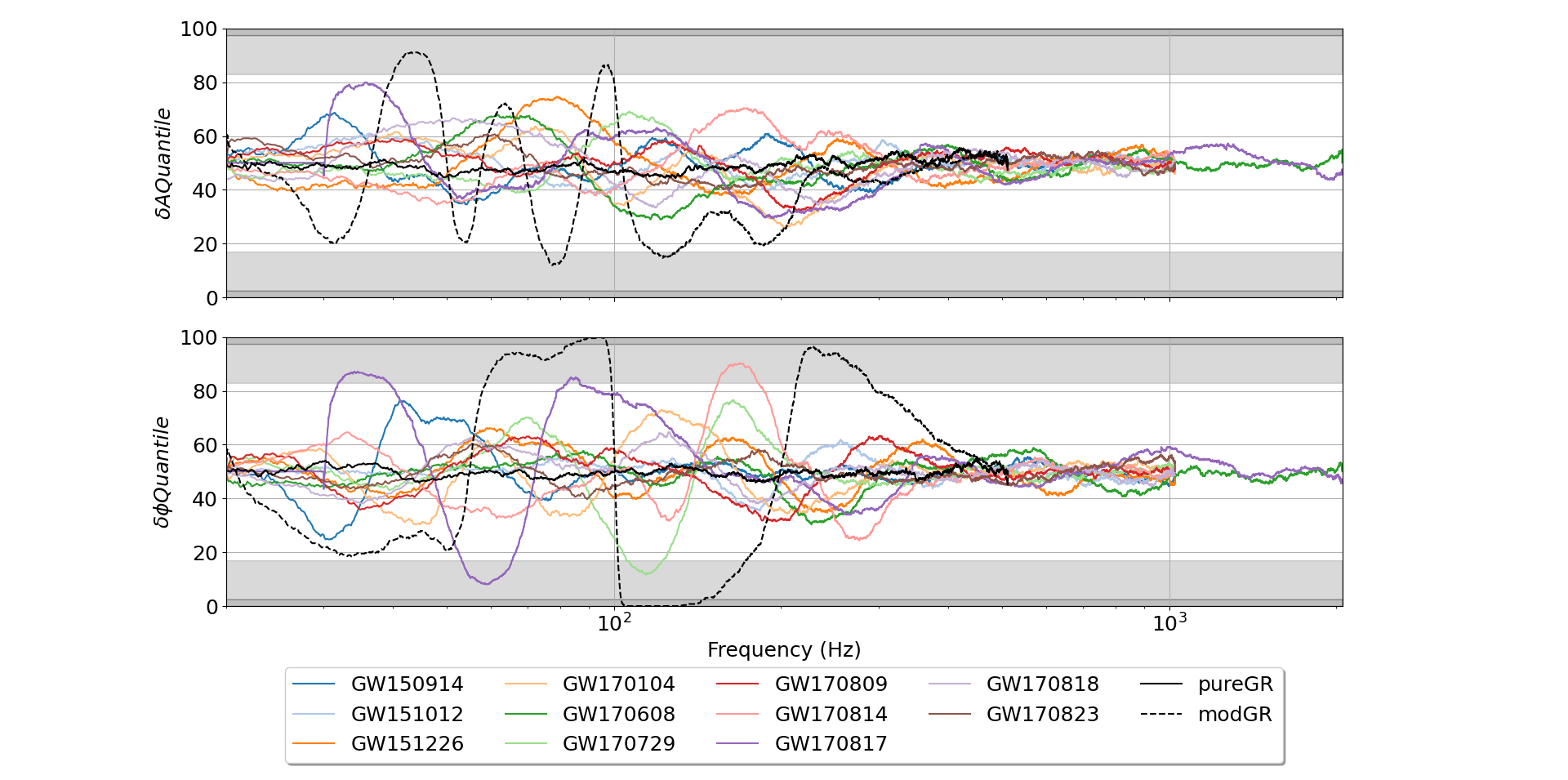}
    \caption{Quantile of spline interpolation that 0-deviation corresponds to for GWTC-1 events. (Different choices of $f_\mathrm{low}$ and sampling rates were chosen for some events.)}
    \label{fig:event_quants}
\end{figure*}{}

The LIGO-Virgo GWTC-1 catalog \cite{catlog_paper} presents a population of GW events to study deviations of GR along with a suite of data with which to validate our gravitational waveform models. We first discuss the results of our model on the ten binary black hole merger events listed with a few selected median posterior parameters in table \ref{table:catalog}, and then we will discuss the one binary neutron star event, GW170817.

We performed parameter estimation on the ten binary black hole (BBH) events with the template waveform IMRPhenomPv2 \cite{IMRPhenomPv2}, and the single binary neutron star (BNS) event, GW170817, with the comparable IMRPhenomPv2\_NRTidal \cite{Dietrich_2019} template waveform that includes neutron star tidal effects, each with our spline model turned on and off. For GW170817, since there was an Electromagnetic Counterpart detected, and to help optimize the speed of sampling, we have fixed the sky location of the source in our analysis. These runs used the same prior settings detailed in Section \ref{sec:spline} which is a 5\% uncertainty of the amplitude and 5 degrees uncertainty on phase. First we look at the spline plot recoveries for GW170823 and GW170729 to show two different cases, both still consistent with zero deviations. In figure \ref{fig:spline_gw170823} we see the case where our model is consistent with zero deviations consistently across the frequency band with the credible intervals illustrating the prior exploration during sampling. Contrasting to this we can look at figure \ref{fig:spline_gw170729} to see a case in which our model has less posterior support for zero deviations at some frequencies. We see here that there is an area on the plot, most importantly the phase portion, around 100 Hz in which zero deviations falls nearly outside our 1$\sigma$ interval. The first thing one might wonder seeing deviations here is whether this can be explained by the normal calibration uncertainties of each detector being similar. We check this in fig \ref{fig:170729_cal_HOM} along with a run using a waveform model including higher-order modes (HOM) or greater than $\ell = 2$ modes in the spherical harmonic expansion \cite{higher-order-modes}, IMRPhenomPv3HM \cite{Khan_2020}. As seen in table \ref{table:catalog} GW170729 is the most massive event from GWTC-1 and HOM are more important for higher mass systems along with asymmetric mass systems \cite{Chatziioannou_2019, Kimball_2020}. We can clearly see a very similar spline interpolant recovery leading us to believe this is fitting features in the data that are unexplained by either a higher-order mode waveform or similar calibration errors across the network of detectors. However for each of the GW170729 results we still have posterior support for zero deviations at the frequencies of largest excursions.

\begin{figure}[h]
    \centering
    \includegraphics[width=0.45\textwidth]{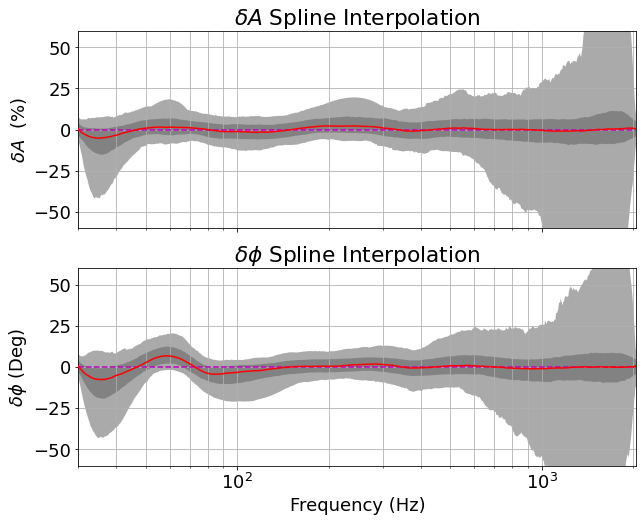}
    \caption{Spline interpolation of GW170817 with 1 and 2 $\sigma$ credible intervals (grey) and the median spline interpolant (red) shown.}
    \label{fig:spline_GW170817}
\end{figure}{}
\begin{figure}
    \centering
    \includegraphics[width=.45\textwidth]{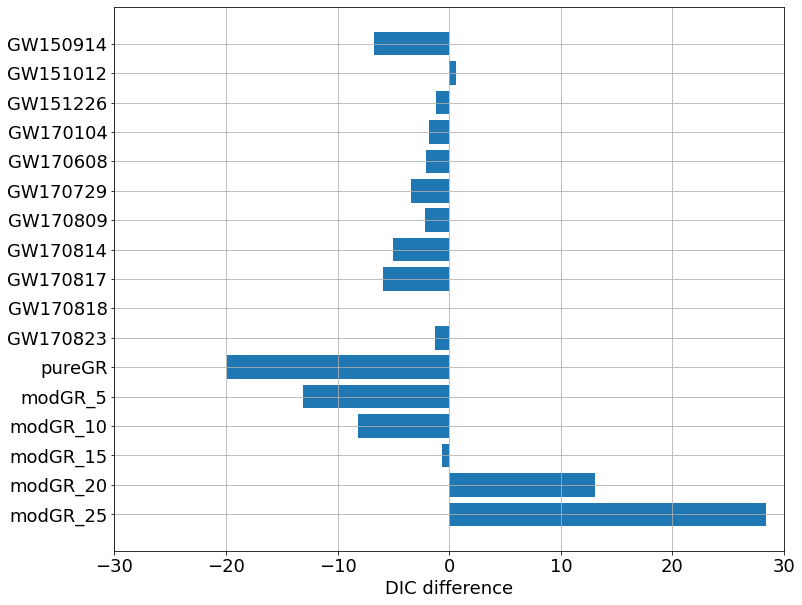}
    \caption{DIC of spline-model-off run minus DIC of spline-model-on run for GWTC-1 events and simulated signals. Simulated modified signals are denoted with the magnitude of phase deviation jump in degrees. Negative values correspond to spline model disfavored while positive values show the spline model favored.}
    \label{fig:DIC_events}
\end{figure}

As in the previous section we can now look at which quantile of the spline interpolant posteriors will fall along the x-axis (corresponding to no deviation), at each frequency bin for each event. This is shown in figure \ref{fig:event_quants} in which we highlight the $>1\sigma$ and $>2\sigma$ bands. We notice in this figure that for all ten binary black hole events, no deviations or modifications in amplitude fall outside the 1-sigma interval. We do see that for two events, namely GW170729 and GW170814, there are regions in which no deviation falls outside of the $1\sigma$ band. However there are significant portions where GW170817 also falls outside the $1\sigma$ band. Looking at the spline posterior in figure \ref{fig:spline_GW170817} for GW170817 shows the clear departures away from zero at some frequencies but overall outside of that small region looks behaved.

To further check our analysis we perform the same model comparisons described earlier in this paper on an event by event basis comparing the spline model on to the spline model off. This is seen in figure \ref{fig:DIC_events} with the same results from differing the phase jump as discussed in the previous section. This shows that for each event there is no preference to either model from their DIC values. Even for the events with more extreme spline interpolant posterior (i.e, GW170729 and GW170817) we still see that from a model comparison approach the spline model does not significantly describe the data better than without the spline model. 


\section{\label{sec:conclusions} Conclusions}

We have presented a useful model and parameterization to describe general departures or deviations from gravitational waveform models. Our model can be used to look for departures from any of modeled waveforms by generically fitting the entire frequency band at once  with spline functions. We find that for the 11 gravitational wave events of both BBH and BNS origin in GWTC-1, the data are consistent with the IMRPhenomPv2 waveform. Shown in figure \ref{fig:event_quants} there are two events that we consider ``outliers'' with some portion of the phase deviation outside of the $1\sigma$ range but still lying well within the $2\sigma$ bounds of no deviations, which for a sample size of 11 events would not be unexpected, even with no deviations present.

Currently, more investigation into possible degeneracies of our model would be necessary to vet any significant sign of deviation. Further studies also need to be done to evaluate effects of detector sensitivity on our model, expand the validation of our model on other physically motivated deviations that can be simulated, and possibly incorporating information from proposed alternatives to GR into the priors. However, the model presented in this paper can be used as a model agnostic test to look for first signs of departures in the modeled waveforms. With more events and increased detector sensitivity, we will be able to better constrain any general deviations with our model while at the same time giving us a better testing set to look for hidden degeneracies between our model and other parameters.

\section{Acknowledgments}

This material is based upon work supported by the National Science Foundation under Grant No. 1807046 and work supported by NSF’s LIGO Laboratory which is a major facility fully funded by the National Science Foundation. We would also like to thank the Niels Bohr Institute for its hospitality while part of this work was completed, and acknowledge the Kavli Foundation and the DNRF for supporting the 2017 Kavli Summer Program. This work made use of the \texttt{PyCBC} \cite{pycbc}, \texttt{GWPy} \cite{GWPY}, and \texttt{LALInference} \cite{LALinference} software packages along with creating all plots and figures with \texttt{Matplotlib} \cite{MATPLOTLIB} and \texttt{corner} \cite{corner}.

\bibliographystyle{apsrev4-1}
\bibliography{bibliography}

\appendix
\section{Degrees of Freedom and Model Comparison}

If deviations from a modeled waveform are primarily in phase but not the amplitude of the signal, our model's added flexibility could unnecessarily penalize our model comparisons in the effective dimensionality penalty of the DIC. The DIC test statistic is defined as:

\begin{multline}\label{dicdef}
    DIC = -2 \overline{\log(\mathcal{L})} + p_{D} \\
    = -2 \left( \overline{\log(\mathcal{L})} - \mathrm{var}(\log(\mathcal{L})) \right)
\end{multline}

With $\overline{\log\mathcal{L}}$ the average log-likelehood, and $p_D$ the effective number of dimensions, defined as $p_D = \frac{1}{2}\mathrm{var}(-2\log\mathcal{L})$ with $\mathrm{var}(...)$ denoting the variance. In this definition the effective dimension term penalizes models with more degrees of freedom. Here we explore the constraints of a phase-only modified model. We only do so in an approximate way, assuming that the goodness of fit (i.e., mean of log-likelihood) term obtained with our full phase and amplitude model is the same that a phase-only model would produce. We then halve the model's effective dimension, $p_D$, in the DIC calculation shown in \ref{dicdef}, effectively removing the degrees of freedom due to allowing the amplitude to vary. 

\begin{figure}[b]
    \centering
    \includegraphics[width=0.5\textwidth, height=0.3\textheight]{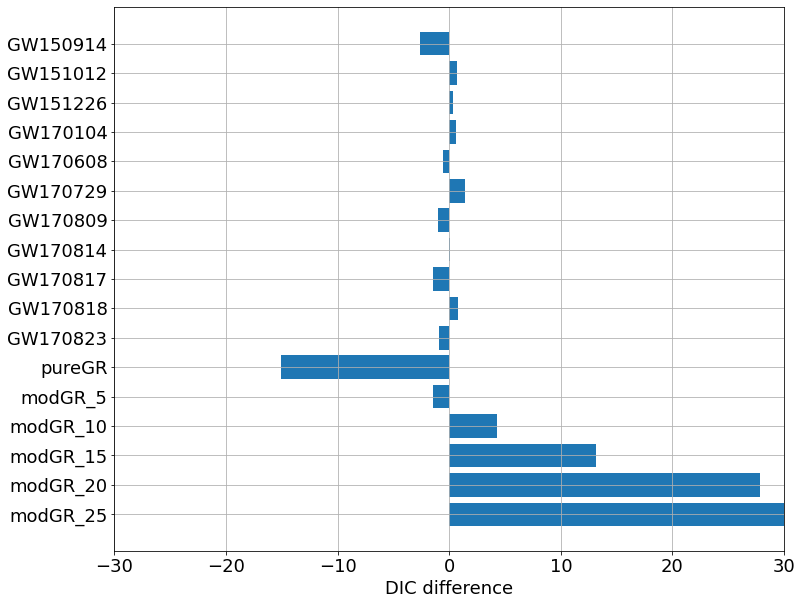}
    \caption{DIC of spline-model-off run minus DIC of spline-model-on run for GWTC-1 events and simulated signals. DICs in this plot are calculated with half of the variance degree of freedom penalty term. Simulated modified signals are denoted with the magnitude of phase deviation jump in degrees. Negative values correspond to spline model disfavored while positive values show the spline model favored.}
    \label{fig:dichalf}
\end{figure}

Figure \ref{fig:dichalf} shows that when approximating a phase-only deviation model as described above, we still do not show a clear favoring of the spline model for any event in GWTC-1, but compared to figure  \ref{fig:DIC_events} there is more ambiguity about which model is favored. This illustrates that with the current data, a phase-only spline deviation model as presented does not qualitatively alter our conclusions but may be useful in analyzing future catalogs of GW events. 

\end{document}